\documentclass[pre,aps,twocolumn,nofootinbib,showpacs,floatfix]{revtex4}
\usepackage{graphics} 
\usepackage{amssymb,amsmath}
\usepackage{amsmath}
\usepackage{psfrag}
\usepackage{graphicx}
\usepackage{subfigure}

\def\be{\begin{equation}} 
\def\ee{\end{equation}} 
\def\bea{\begin{eqnarray}} 
\def\eea{\end{eqnarray}}  
\def\bean{\begin{eqnarray*}} 
\def\eean{\end{eqnarray*}} 
\def\dd{\partial}

\def\bv{{\bf v}}
\def\br{{\bf r}}

\def\bse{\begin{subequations}}
\def\ese{\end{subequations}}

\def\bF{{\mathbf F}}

\def\bV{{\mathbf V}}

\def\lsim{\raise 0.4ex\hbox{$<$}\kern -0.8em\lower 0.62ex\hbox{$\sim$}} 
\def\gsim{\raise 0.4ex\hbox{$>$}\kern -0.7em\lower 0.62ex\hbox{$\sim$}}

\def\f0N{f_0^{(N)}}
\def\bec{\begin{center}}
\def\eec{\end{center}}

\begin{document} 

\title{Collisional relaxation of two-dimensional self-gravitating systems}

\author{B.~Marcos} 
\email{bruno.marcos@unice.fr}

\affiliation{Laboratoire J.-A. Dieudonn\'e, UMR 6621,
Universit\'e de Nice --- Sophia Antipolis,\\
Parc Valrose 06108 Nice Cedex 02, France}

\begin{abstract}   
Systems with long range interactions present generically the formation
of quasi-stationary long-lived non-equilibrium states. These states
relax to Boltzmann equilibrium following a dynamics which is not well
understood. In this paper we study this process in two-dimensional
inhomogeneous self-gravitating systems.  Using the Chandrasekhar -- or
local -- approximation we write a simple approximate kinetic equation
for the relaxation process, obtaining a Fokker -- Planck equation for
the velocity distribution with explicit analytical diffusion
coefficients. Performing molecular dynamics simulations and comparing
them with the evolution predicted by the Fokker -- Planck equation, we
observe a good agreement with the model for all the duration of the
relaxation, from the formation of the quasi-stationary state to
thermal equilibrium. We observe however an overestimate or
underestimate of the relaxation rate of the particles with the slower
or larger velocities respectively. It is due to systematic errors in
estimating the velocities of the particles at the moment of the
collisions, inherent to the Chandrasekhar approximation when applied to
inhomogeneous systems.  Theory and simulations give a scaling of the
relaxation time proportional to the number of particles in the system.
\end{abstract}    
\pacs{ 04.40.-b, 05.70.Ln, 05.70.-a}    
\maketitle   
\date{today}  

\twocolumngrid

\section{Introduction}

Systems of particles with long range interactions are those which
inter-particle potential at large separation decays slower than the
dimension $d$ of space, i.e., $v(r\to\infty)\sim1/r^\gamma$ with
$\gamma \le d$. There are many examples in nature, such as
self-gravitating systems in the cosmological and astrophysical context
(the large structure of the universe, galaxies, etc), interaction
between vortices in two-dimensional hydrodynamics, cold classical
atoms or capillary interactions between colloids or granular media (for
a review see e.g.  \cite{campa_etal_LRreview_2009}). These kinds of
systems present very particular properties in thermal equilibrium,
such that negative micro-canonical specific heat or inequivalence of
statistical ensembles. Their dynamics is also peculiar compared to
short range systems: in a first stage there is the generic formation
in a few characteristic times $\tau_{dyn}$ of a long-lived non-equilibrium
state --- during the so-called \emph{violent relaxation} process. A
typical example of such quasi-stationary states (hereafter QSS) are
galaxies or young globular clusters. Then, a comparatively very slow
relaxation to thermodynamical equilibrium occurs --- called
\emph{collisional relaxation} --- in a timescale of order
$\tau_{coll}\sim N^\delta \tau_{dyn}$, where $N$ is the number of
particles and $\delta\ge 1$ depends on the system studied.

The mechanism of collisional relaxation is still not well
understood. In the context of gravitational systems, Chandrasekhar
found theoretically, in a seminal work \cite{chandra_42}, an estimate
of the relaxation time for gravitational systems in three
dimensions. He considered an homogeneous system and computed the
change in velocity due to successive independent
collisions\footnote{We will use here, as in the astrophysical
  literature, the term ``collisions''. In the general context of
  long-range systems it would be more appropriate to call them
  ``finite $N$ effects''.} 
of a test particle in a stationary
macroscopic configuration.  Because of the hypothesis of homogeneity
there is no macroscopic scale in the system, which led to an ongoing
controversy about the value of the maximal impact parameter of the
collisions and in particular how it should scale with $N$
\cite{henon_58,farouki_82,smith_92,farouki_94}. Following this,
several studies considered collective effects
(e.g. \cite{weinberg_93}), but still in homogeneous configurations. An
explicit theoretical description of the collisional relaxation in
inhomogeneous systems is technically much more difficult to derive,
being necessary the use of action-angle variables. This description is
still lacking, despite recent progress in this direction
\cite{heyvaerts_10,chavanis_12b}, for a recent review see
e.g. \cite{chavanis_12c}. 

The collisional relaxation has also been studied numerically, for a
wide variety of systems. For one-dimensional
gravity, a scaling of $\tau_{coll}\sim N \tau_{dyn}$ has been measured
for the full relaxation process \cite{joyce_10}, and in the
Hamiltonian Mean Field model the scaling has been found to be
dependent on the initial condition: $\tau_{coll}\sim N \tau_{dyn}$
\cite{yamaguchi_etal_04}, $\tau_{coll}\sim N^{1.7} \tau_{dyn}$
\cite{yamaguchi_etal_04} or $\tau_{coll}\sim \exp(N) \tau_{dyn}$
\cite{campa_08}. For dimensions larger than $d=1$, the relaxation has
been estimated studying --- for numerical reasons --- only its early
stage, i.e., for times in which the QSS is weakly perturbed (see
e.g. \cite{diemand_04,gabrielli_10}), or performing simulations with a
simplified dynamics. For gravity in two-dimensions, in simulations
performed imposing radial symmetry, it has been observed
$\tau_{coll}\sim N^{1.35} \tau_{dyn}$ \cite{teles_10}. In $d=3$
dimensions, the Chandrasekhar scaling $\tau_{coll}\sim N/\ln N
\tau_{dyn}$ has been verified for gravity
(e.g. \cite{farouki_82,diemand_04,marcos_13}) and for power-law
potential $u(r)=1/r^\gamma$, for which has been found $\tau_{coll}\sim
N \tau_{dyn}$ if $\gamma < 2$, see \cite{gabrielli_10,marcos_13}.

In this paper, we study the collisional relaxation of a
self-gravitating system in $d=2$ dimensions.  The interacting
potential --- solution of the Poisson equation in $d=2$ dimensions ---
is $u(r)=g\ln(r)$, where $g$ is the coupling constant. It is an
attractive model because it presents the same mechanism of collisions as in
$d=3$ (which is not the case for models in $d=1$), the system is
self-confined (it is not necessary to confine it artificially in a
box), thermal equilibrium properties are easily calculated and
numerical simulations are easier to perform than in $d=3$.  Moreover,
as mentioned above, it was found in \cite{teles_10}, using simulations
imposing the radial symmetry (particles conserve their initial angular
momentum), that the collisional relaxation scales with the number of
particles in the unexpected manner $\tau_{coll}\sim N^{1.35}
\tau_{dyn}$. In the way in which these simulations have been performed
the actual model is quasi one-dimensional, and this result may be in
some connection with the striking relaxation time for the HMF model,
in which for some initial conditions it has been found to scale as
$\tau_{coll}\sim N^{1.7} \tau_{dyn}$. 

Another question that will be addressed in this paper is the fact that
it has been observed that the Chandrasekhar approximation --- or \emph{local approximation} --- gives good
estimation of the relaxation time not only in homogeneous systems but
also in non-homogeneous configurations (see
\cite{,farouki_82,smith_92,farouki_94,gabrielli_10}), and in
particular how it scales (in a non-trivial way) with the number of
particles $N$ and the minimal impact
parameter \cite{gabrielli_10}. This suggests the possibility to
describe, in a good approximation, the whole collisional relaxation process
using this approximation (see e.g. \cite{chavanis_12c}), in which
the system is treated as locally homogeneous.

This paper is organized as follows. In the Sect.~\ref{theory}, we
show that, if the QSS which is collisionally relaxing is approximately
homogeneous in its center --- as it is for many initial conditions for
gravitational system in $d=2$ and $d=3$ --- then treating the system
as homogeneous (but finite) is a reasonable approximation. Then, we
compute the diffusion coefficients and, neglecting collective effects,
we write a Fokker -- Planck equation which describes the evolution of
the system. In Sect.~\ref{simulations}, we report simulations using
molecular dynamics of the relaxation of the system, for the whole time
range between the QSS and the final thermal equilibrium, for two
different initial conditions and different number of particles. We
will see, that despite the many approximations, the evolution of the
velocity pdf is reasonably well described by the theory for
intermediate values of the velocity. In Sect.~\ref{sect-chandra} will
discuss the validity of the Chandrasekhar approximation.  In
Sect.~\ref{conclusion} we present the conclusions of this study and
further perspectives.

\section{Theoretical description}
\label{theory}
 We model the generic evolution of the system using the Boltzmann equation for the one point probability density function $f(\br,\bv,t)$. We can write it formally as
\be
\label{VFP}
\frac{\dd f}{\dd t} + \bv\cdot\frac{\dd f}{\dd \br} + \bF[f]\cdot \frac{\dd f}{\dd \bv} = \Gamma_c[f],
\ee
where $\Gamma_c[f]$ is the collision operator. During the relaxation
process, the system reaches first a QSS and then evolves
(comparatively slowly) through an infinity sequence of QSS, in which
\be
\label{aprox}
\bv\cdot\frac{\dd f}{\dd \br}+\bF[f]\cdot \frac{\dd f}{\dd
  \bv}=0.
\ee
To make Eq.~\eqref{VFP} tractable analytically, we will assume that Eq.~\eqref{aprox} holds for all times, which implies not taking collective effects
into account.

We will focus in this paper on the evolution of the velocity pdf
\be
s(\bv,t)=\int d^2r\,f(\br,\bv,t).
\ee
 We integrate Eq.~\eqref{VFP} over
the positions, obtaining, in the approximation \eqref{aprox}
\be
\label{VFP-vel}
\frac{\dd s}{\dd t} = \int d^2r\,\Gamma_c[f].  \ee
In the same manner as in the most studied $d=3$
case, the relaxation is dominated by \emph{weak collisions} (see
e.g. \cite{binney}), i.e., the ones for which the trajectories of the particles are weakly perturbed. Moreover,
it has been shown that, for times larger than one orbital period, the
force correlation function decays rapidly (e.g. as $\sim 1/t^5$ for
gravity in $d=3$ \cite{cohen_75a}). We may then consider that
collisions are independent and the use of a Fokker-Planck
approximation of Eq.~\eqref{VFP-vel} is therefore justified (see
e.g. \cite{chavanis_12c,risken_89}), which can be written as
\be
\label{VFP-simp}
\frac{\dd s(\bv,t)}{\dd t} = \frac{\dd}{\dd v_i}\left[D_{v_i}s(\bv,t)\right]
+ \frac{1}{2} \frac{\dd^2}{\dd v_i \dd v_j}\left[D_{v_i v_j}s(\bv,t)\right],
\ee
where the diffusion coefficients are defined as average change of the velocity of
the particles per unit of time, i.e.,
\bse
\label{diff-coeff}
\begin{align}
D_{v_i}(\bv) &= \frac{\langle \Delta v_i\rangle}{\Delta t}\\
D_{v_iv_j}(\bv) &= \frac{\langle\Delta v_i \Delta v_j\rangle}{\Delta t}.
\end{align}
\ese
In Eqs.~\eqref{VFP-simp} and \eqref{diff-coeff} we have assumed that
the diffusion coefficients are a well defined quantity to describe the
relaxation process in an inhomogeneous system. We will see in what
follows to what extent it is a good approximation.

The strategy to compute the diffusion coefficients is the following: because collisional relaxation is dominated by weak collisions, i.e.,
by the ones in which the trajectory of the particles are weakly
perturbed (see e.g. \cite{binney}), the diffusion coefficients \eqref{diff-coeff} can be calculated computing changes in velocity of the particles considering that they are evolving on their unperturbed orbits (i.e. the ones which corresponds to the mean field $N\to\infty$ limit). In Subsect.~\ref{mean-field-potential} we will then first estimate the mean mean potential in which the particles are evolving , in Subsect.~\ref{comp-change-vel} we will then compute the change in velocity due to one collision  and finally in Subsect.~\ref{comp-diff-coeff} we will compute the diffusion coefficients themselves.

\subsection{Mean field potential}
\label{mean-field-potential}

\begin{figure}
  \begin{center}
        {\includegraphics[height= 0.35\textwidth]{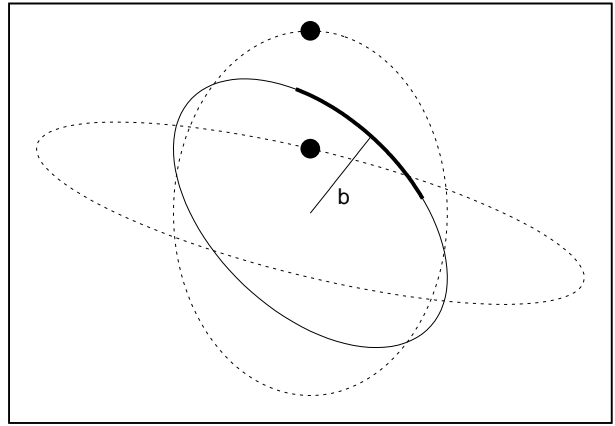}}
\end{center}
\caption{Sketch of the orbits (dotted curves) of two ``colliding''
  particles (which are plotted at the same arbitrary time). The plain
  curve represents their relative trajectory, and the thick portion
  (of length $\sim 2b$) the part of the trajectory in which $|\Delta
  \bV_\perp|$ changes significantly (see text).}
\label{orbit}
\end{figure}

 We are going to assume that in the region in which particles are
 collisionally relaxing the density pdf is homogeneous. This
 distribution generates a harmonic gravitational field. We will see in
 our simulations (see Sect.~\ref{simulations}) that it is a very good
 approximation. Moreover, the this is also true for the thermal
 equilibrium state, which is the final state the system will reach. At
 thermal equilibrium the potential generated by the QSS (see
 e.g. \cite{teles_10})
\be
\label{pot_MB}
\Psi(r)=\frac{gN}{2}\ln\left(\lambda^2+r^2\right)
\ee
 where $\lambda$ is a constant which depends on the total energy of
 the system\footnote{The $N\to\infty$ limit is taken in such a way
   that $g\propto N^{-1}$, which is equivalent to keep the dynamical
   time of the system invariant changing $N$, see
   Eq.~\eqref{tau_dyn}. We keep here the dependence on $N$ to have an
   explicit dependence on $\tau_{dyn}$ in our equations.}. For
 $r\lesssim\lambda$ (which corresponds to a scale in which are
 included half of the particles), the potential is harmonic, i.e.,
\be
\label{pot_MB_taylor}
\Psi(r)\simeq gN\ln \lambda+ \omega^2 r^2,
\ee
where 
\be
\omega^2=\frac{gN}{2\lambda^2}.
\ee
 Under the  hypothesis that the potential has the form \eqref{pot_MB_taylor}, the trajectories of the particles in the
central region of the system (where collisional relaxation occurs) can
be then well approximated with ellipses. The relative motion of two
particles is also therefore an ellipse which can be written as
\be
\label{elip}
\br(t)=x_0\sin(\omega t)\hat x + y_0\cos(\omega t)\hat y,
\ee
as it has been sketched in Fig.~\ref{orbit}. We expect that the
hypothesis \eqref{pot_MB_taylor} is relatively general: it has been
shown numerically in $d=3$ that, for a wide set of initial
conditions, the QSS present also a central homogeneous region which
decays rapidly to zero at larger scales \cite{roy+perez_2004}.

\subsection{Computation of the change of velocity due to one ``collision''}
\label{comp-change-vel}

\begin{figure}
  \begin{center}
    \psfrag{X}[][]{$y_0/x_0$}
    \psfrag{Y}[][]{$|\Delta \bV_y|w x_0/g\pi$}
    \includegraphics[height= 0.35\textwidth]{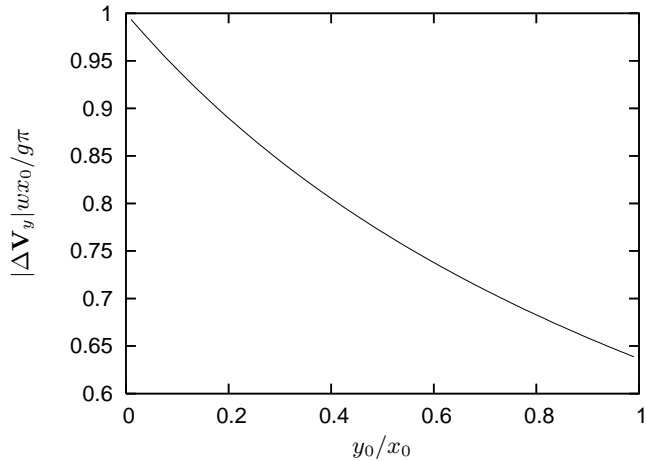}
\end{center}
\caption{Continuous line: change in the relative velocity in the $y$ direction $|\Delta \bV_y|$ Eq.~\eqref{delta_vy} normalized by its asymptotic value \eqref{delta_approx} as a function of the ellipticity  $y_0/x_0$.}
\label{fig-approx}
\end{figure}

In the context of long range
systems, we define a ``collision'' between two particles as the
process in which they cross each other in half an orbital period (one
crossing of the system). Assuming that the
relative orbits have the form of Eq.~\eqref{elip} we can compute the
change in relative velocity in the $\hat y$ direction of two crossing
particles by integrating the gravitational acceleration $\bF(t)/m$ projected in the $\hat y$ direction over the duration of a collision:
\bea
\label{delta_vy}
\nonumber
|\Delta \bV_y|&=&2g\int_0^\frac{\pi}{2\omega} \frac{\bF(t)\cdot \hat y}{m}dt\\\nonumber
&\simeq& 2g \int_0^\frac{\pi}{2\omega}\frac{y_0\cos(\omega t)\,dt}{x_0^2 \sin^2(\omega t) + y_0^2 \cos^2(\omega t)}\\
&=& 2g\frac{\arctan\left[\sqrt{\frac{x_0^2}{y_0^2}-1}\right]}{w \sqrt{x_0^2-y_0^2}}.
\eea
From geometrical arguments, it is possible to see that most
of the orbits will have large ellipticity. For example, in our simulations we find  $y_0/x_0\approx 0.1$ on average (see Sect.~\ref{sect-chandra}). If we choose the axis in order $y_0 < x_0$, then, if the condition 
\be
\label{large_elip}
y_0\ll x_0,
\ee
holds, Eq.~\eqref{delta_vy} can be well approximated by
\be
\label{delta_approx}
|\Delta \bV_y|= \frac{g\pi}{\omega x_0}\left(1+\mathcal O\left(\frac{y_0}{x_0}\right)\right).
\ee
In  Fig.~\ref{fig-approx} we show how the approximation \eqref{delta_approx} becomes better increasing the ellipticity $x_0/y_0$. For example, a maximal relative error of $35\%$ is made for $x_0/y_0 = 1$ decreasing rapidly to an error of $6\%$ when $x_0/y_0 = 0.1$.
From Eq.~\eqref{delta_vy} it is possible to see that the ``collision''
is localized in space and time: as the integral converges rapidly, an
excellent approximation of \eqref{delta_vy} --- with the condition
\eqref{large_elip} --- consists in taking as upper cutoff of the
integral $\omega t\simeq y_0/x_0$. This means that most of the change
of velocity occurs during the interval of time $\Delta t \simeq \omega^{-1}
y_0/x_0$ centered around $t=0$ in our parametrization \eqref{elip},
in a region of length $\sim 2y_0$.

In order to compute simply averages over the velocity pdf in what follows, it is useful to have an expression of the
change of velocity as a function of the velocity of the particle
itself. In the same approximation \eqref{large_elip} we have
\be
\label{v-approx}
|\bV(t=0)|\equiv V\simeq \omega x_0\left(1+\mathcal O\left(\frac{y_0}{x_0}\right)\right).
\ee
Then
\be
\label{v_perp}
|\Delta \bV_y|\equiv|\Delta\bV_\perp|\simeq \frac{g\pi}{V},
\ee
where $V$ is the relative velocity at the distance of closest
approach. We use the notation $\bV_\perp$ because, in this
approximation, $\Delta \bV_y$ corresponds to the change of velocity in
the perpendicular direction of the velocity of the particle. This
result is the one obtained by Chandrasekhar adapted to
self-gravitating systems in $d=2$ dimensions. We will discuss the
implications and limitations of this approach in
Sect.~\ref{sect-chandra}.

It is possible to compute the change in the relative
parallel velocity using that, in a weak collision, $V$ does not change during the collision. Then:
\bse
\begin{align}
|\Delta
\bV_\perp|&=V \sin\theta\\
|\Delta \bV_\parallel|&=V(1-\cos\theta),
\end{align}
\ese
where $\theta$ is the angle of deflection. In the weak collision
approximation $\theta\ll 1$ and thus we have $\sin\theta\simeq\theta$
and $\cos\theta\simeq 1-\theta^2/2$ and then
\be
\label{v_par_perp}
|\Delta \bV_\parallel|= \frac{|\Delta \bV_\perp|^2}{2V}.
\ee
Taking into account that particle masses are equal we obtain for the
change in velocity of a particle, using Eqs.~\eqref{v_perp} and
\eqref{v_par_perp}, 
\bse
\begin{align}
|\Delta \bv_\perp|&\simeq \frac{\pi g}{2V}\\
|\Delta \bv_\parallel|&\simeq \frac{\pi^2 g^2}{4 V^3}.
\end{align}
\ese

\subsection{Computation of the diffusion coefficients}
\label{comp-diff-coeff}
We compute the diffusion coefficients using the standard method used
in $d=3$ in the local approximation.  As the spatial density
pdf is approximately constant up to a scale $r^*$ in radial
coordinates (see discussion above and numerical simulations of
Sect.~\ref{simulations}), we can therefore estimate the number $\eta$
of collisions of a particle in an time interval $\Delta t$, on average, as
\be
\label{coll}
\eta\simeq \frac{2 NV\Delta t}{\pi r^*};
\ee
the factor $\pi r^*/2$ is the average height of a circle of radius
$r^*$. We are going now to average over the velocity pdf. We will do a
somewhat uncontrolled approximation here because Eq.~\eqref{v_perp}
gives the change of relative velocity \emph{at the point of closest
  approach}. It is not possible to compute exactly this quantity from
the velocity pdf because the change in velocity of a particle does not
depend on its velocity (as in the homogeneous case) but in the orbit
to which it behaves, i.e., in the particular values of $x_0$ and $y_0$
corresponding to the particle. To go further, however, we will assume
that it is possible to average over the velocity pdf $s(v)$. 
Introducing, as in the $d=3$ case, the Rosenbluth potential
\cite{rostoker_60}
\bse
\begin{align}
q(v) &= \int d^2v' \frac{s(v')}{|\bv-\bv'|}\\
p(v) &= \int d^2v' s(v')|\bv-\bv'|,
\end{align}
\ese
and  assuming that the velocity pdf is isotropic, we obtain, keeping only terms of $\mathcal O(g^2)$ (see
App.~\ref{derivation}):
\bse
\begin{align}
D_{v_i}(v)&=\frac{\langle\Delta v_i\rangle}{\Delta t}=C\frac{\dd q(v)}{\dd v_i}\\\nonumber
D_{v_iv_j}(v) &=\frac{\langle\Delta v_i \Delta v_j\rangle}{\Delta t}=C\frac{\dd^2 p(v)}{\dd v_i \dd v_j},
\end{align}
\ese
where 
\be
\label{coeff}
C=\frac{\pi g^2N}{2 r^*}.
\ee
As the succession of QSS have an approximate polar symmetry, it is then
useful to write Eq.~\eqref{VFP-simp} in polar coordinates. Considering that
the Rosenbluth potentials are isotropic, we get using Eq.\eqref{rosen-polar}
\be
\label{VFP-polar-pre}
\frac{\dd \tilde s}{\dd t} = C \left\{-\frac{\dd }{\dd
  v}\left[\left(q'( v)+\frac{p'( v)}{2 v^2}\right)\tilde
  s\right]+\frac{1}{2}\frac{\dd^2 }{\dd v^2}\left[ p''( v)\tilde
  s\right]\right\}, \ee
where $\tilde s(v)$ as the velocity pdf in polar coordinates,
$v=|\bv|$ and the primes denotes derivation with respect to $v$. It is
useful to write Eq.~\eqref{VFP-polar-pre} in an adimensional form. We
define the our time unit as the dynamical time of the system
\be
\label{tau_dyn}
\tau_{dyn}=\frac{1}{\sqrt{gN}}.
\ee
 We define the velocity units $v_*$ using the
virial theorem, which states that, for any stationary state (and hence
a QSS), the average velocity square of the particles is constant
during the evolution (e.g. \cite{chavanis_06,teles_10}):
\be
\label{virial}
\langle v^2\rangle = \frac{g N}{2}.
\ee
 It is then natural to take as
velocity unit 
\be
v_*=\sqrt{gN}.
\ee
Defining the adimensional time and
velocities as $\tilde t = t/\tau_{dyn}$ and $\tilde v= v/v_*$ respectively, we get from Eq.~\eqref{VFP-polar-pre}
\be
\label{VFP-polar}
\frac{\dd \tilde s}{\dd \hat t} = \hat C \left\{-\frac{\dd }{\dd \hat v}\left[\left(q'(\hat v)+\frac{p'(\hat v)}{2\hat v^2}\right)\tilde s\right]+\frac{1}{2}\frac{\dd^2 }{\dd \hat v^2}\left[ p''(\hat v)\tilde s\right]\right\},
\ee
where we have defined
\be
\label{hat_C}
\hat C = C\frac{\tau_{dyn}}{v_*^3}=\frac{\pi}{2 N r_*}.
\ee
Equation \eqref{VFP-polar} depends on $N$ through  $\hat C$,
which implies that the relaxation scales as
\be \tau_{coll}\sim N \tau_{dyn}.  \ee To compute explicitly the
diffusion coefficients we need an explicit form of $\tilde s(\hat
v)$. As discussed above, the velocity pdf at the
distance of closest approach is unknown. We will use then the standard approximation  to take the equilibrium  Maxwell -- Boltzmann pdf (see e.g. \cite{chavanis_12c})
\be
\label{distri_MB}
\tilde s_{MB}(\hat v)=2\hat v v_*\beta\exp(-\beta \hat v^2),
\ee
with $\beta=2$ given by Eq.~\eqref{virial}. We obtain in this
approximation
\bse
\label{coeff-expl-bessel}
\begin{align}
q(\hat v)&=e^{-\beta \hat v^2/2} \sqrt{\pi \beta}\,I_0\left(\frac{\beta \hat v^2}{2}\right)\\\nonumber
p(\hat v)&=\frac{1}{2}\sqrt{\frac{\pi }{\beta}} e^{-\beta \hat v^2/2} \left[-e^{\beta \hat v^2/2}+\left(1+\beta \hat v^2\right)I_0\left(\frac{\beta \hat v^2}{2}\right)\right.\\
&\,\,\,+\beta \hat v^2 I_1\left(\frac{\beta \hat v^2}{2}\right)\Big],
\end{align}
\ese
where $I_n(x)$ is the modified Bessel function of the first kind. It
is possible to verify that the equilibrium pdf
\eqref{distri_MB} is a stationary solution of Eq.~\eqref{VFP-polar}
with the diffusion coefficients given by
Eq.~\eqref{coeff-expl-bessel}. Note that we obtain the same result
obtained in \cite{chavanis_06b} (see also \cite{chavanis_12d}), in
which a different method to compute the diffusion coefficients
than Rosenbluth potentials has been used.

\section{Numerical simulations} 
\label{simulations}

\begin{figure}
  \begin{center}
    \psfrag{A}[r][r]{$\mu_0=1$}
    \psfrag{B}[r][r]{$\mu_0=1.7$}
     \psfrag{X}[c][c]{$r$}
    \psfrag{Y}[][]{$N(r)/(2\pi r)$}
    \includegraphics[height= 0.35\textwidth]{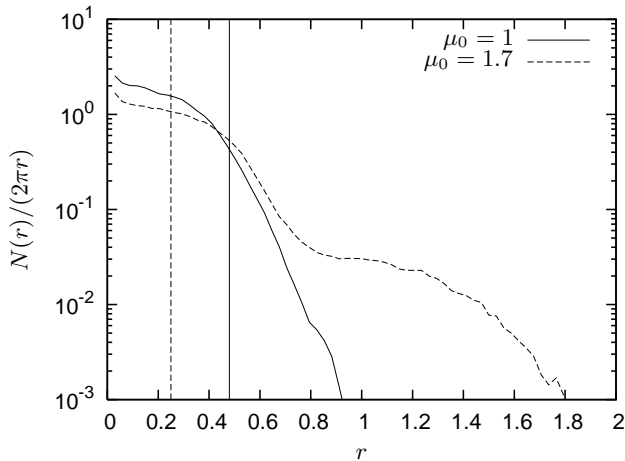}
\end{center}
\caption{Density pdf in the QSS at $t=50\tau_{dyn}$ for both
  initial condition. The vertical curves (of the same type of their
  corresponding density profile) are the values of $r^*$ used in
  \eqref{coll} in order to obtain the measured relaxation rate in the
  simulations.}
\label{density}
\end{figure}

\begin{figure}
  \begin{center}
    \psfrag{A}[r][r]{$t=50\tau_{dyn}$}
    \psfrag{B}[r][r]{$t=5600\tau_{dyn}$}
     \psfrag{X}[][]{$r$}
    \psfrag{Y}[][]{$\Psi(r)-\Psi(0)$}
    {\includegraphics[height= 0.35\textwidth]{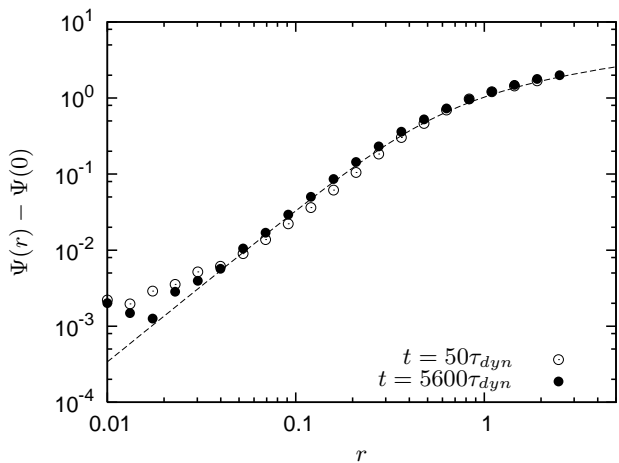}}
\end{center}
\caption{Potential of the particles in function of their radial distance for the simulation with $\mu_0=1.7$, at $t=50\tau_{dyn}$ and $t=5600\tau_{dyn}$. The dashed line is the potential of the distribution at thermal equilibrium \eqref{pot_MB}.}
\label{potential}
\end{figure}

We compare the theoretical model with molecular dynamics simulations
performed with a modification of the publicly available code {\tt
  GADGET2} \cite{gadget} to handle the logarithmic interaction. We use
a time-step of $2.5\times10^{-4}\tau_{dyn}$ in order to ensure a very
precise energy conservation, which is better than $10^{-5}$ for the
whole duration of the runs. We performed simulations with initial
water-bag conditions with different number of particles in the interval
$N=[100,8000]$ and initial virial ratio $\mu_0=1$ and $\mu_0=1.7$, where 
\be
\mu_0=\frac{v_*}{\sqrt{2\langle
  v_0^2\rangle}},
\ee
where $\langle v_0^2\rangle$ is the average of the initial velocity
square. The simulations have been performed for times of
$5600\tau_{dyn}$ for the systems with the largest $N$ and
$7700\tau_{dyn}$ with the smallest one. In order to improve
statistics, we average the measured velocity pdf over $100$
consecutive snapshots in an interval of $2.5 \tau_{dyn}$. The system
forms a QSS which is approximately homogeneous in its central region,
with a rapid decay of the density at larger scale, as it can be seen in
Fig.~\ref{density} for both initial conditions. We observe that the
one with initial virial ratio $\mu_0=1$ gives rise to a compact
density pdf whereas the one with initial virial ratio $\mu_0=1.7$ to a
core halo distribution.
In Fig.~\ref{potential} we plot the potential energy $\Psi(r)$  generated by the
density pdf at time $t=50\tau_{dyn}$ (time in which the system has
violently relaxed) and $t=5600 \tau_{dyn}$, corresponding to thermal
equilibrium for the $\mu_0=1.7$ case (an analogous result is obtained
for $\mu_0=1.7$). We observe that for the inner part of the system the
potential is very well approximated by the potential generated by the
system at thermal equilibrium \eqref{pot_MB_taylor}.
We monitor how
the system approaches thermal equilibrium using the parameter
\be
\xi(t)=\frac{1}{N^2}\int_0^\infty [s(v,t)-s_{MB}(v)]^2dv.
\ee
\begin{figure}
  \begin{center}
    \psfrag{A}[r][r]{$N=16^3$}
    \psfrag{B}[r][r]{$N=12^3$}
    \psfrag{C}[r][r]{$N=750$}
       \psfrag{X}[][]{$t/(0.25 N \tau_{dyn})$}
    \psfrag{Y}{$\xi(t)$}
    {\includegraphics[height= 0.35\textwidth]{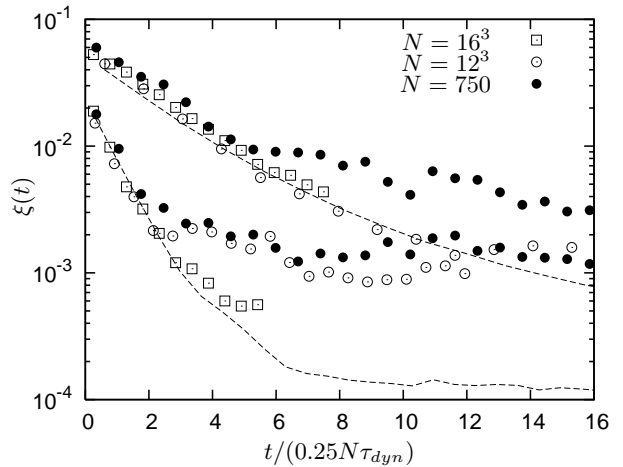}}
\end{center}
\caption{Upper curves: initial condition with $\mu_0=1$. Lower curves: initial conditions with $\mu_0=1.7$. Points: evolution of the crossover parameter $\xi(t)$ measured in the molecular dynamics simulations for the two different initial condition $\mu_0=1$ and $\mu_0=1.7$. Lines: theoretical prediction calculated using Eq.~\eqref{langevin} for each case (see text).}
\label{relaxation_MB}
\end{figure}
In order to compare simulations with theory we compute the associated
Langevin equation of Eq.~\eqref{VFP-polar}. Therefore, the change in
the velocity is given, following the Ito definition, by
\be
\label{langevin}
d\hat v(\hat t)=\hat C\left\{\left(q'(\hat v)+\frac{p'(\hat v)}{2\hat v^2}\right)d\hat t+\sqrt{p''(\hat v)}dW\right\},
\ee
where $dW$ is a Gaussian stochastic variable delta correlated in time
with variance unity. We choose as initial condition a configuration of
the numerical simulation at $t=50\tau_{dyn}$ (time in which the system
has violently relaxed) and then we compare the evolution predicted by
the Langevin equation and the one of the full numerical simulation. We
integrate Eq.~\eqref{langevin} by a simple Euler procedure. In
Fig.~\ref{relaxation_MB} we show the evolution of $\xi(t)$, where the
time axis has been rescaled by a factor $N$, which indicates a scaling
of the relaxation time as $\tau_{coll}\sim N \tau_{dyn}$.  For
clarity, between all the simulations with different numbers of
particles performed we plot three of them. The part of the curve which
flattens corresponds to thermal equilibrium, which is attained first
as $N$ decreases. The matching between the curves corresponding to
different $N$ is very good in the region out of equilibrium, as it has
been illustrated for $N=750$, $N=12^3$ and $N=16^3$, which confirm the
prediction of Eq.~\eqref{VFP-polar} for the scaling of the relaxation.
The dashed curves corresponds to the theoretical prediction given by
Eq.~\eqref{langevin} with $r^*=0.48$ for the simulation with $\mu_0=1$
and $r^*=0.25$ for the simulation with $\mu_0=1.7$. These values are,
within a factor $2$, close to the scale of the falloff in the density
pdf; the density decays to half its center value around $r\approx 0.4$
for both set of simulations. We emphasize that the difference in the
slopes of the curves is essentially due to the different initial
conditions considered for each case rather than in the value of $r^*$
taken: taking indeed the same value of $r^*$ for both initial
conditions the two curves appear to be very different. The full
simulation curves decay to a lower value at thermal equilibrium
because fluctuations appears to be larger in the molecular dynamics
simulations than in the Langevin simulation.
\begin{figure*}[]
  \begin{center}
    \psfrag{X}[][]{$v/v_*$}
    \psfrag{Y}[][]{$s(v)/v$}
    {\includegraphics[height= 0.17\textwidth]{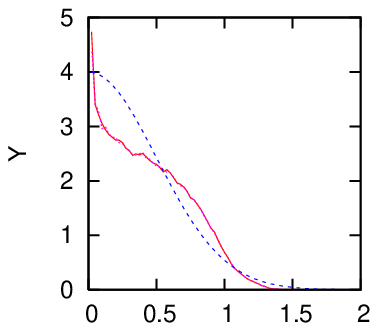}}
    \hspace{-0.3cm}
    {\includegraphics[height= 0.17\textwidth]{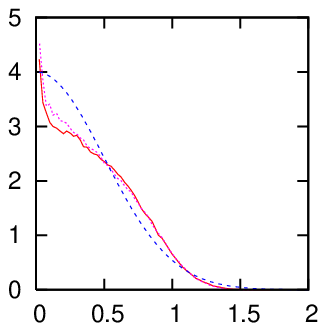}}
    \hspace{-0.3cm}
    {\includegraphics[height= 0.17\textwidth]{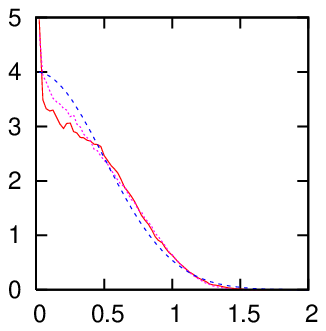}}
    \hspace{-0.3cm}
    {\includegraphics[height= 0.17\textwidth]{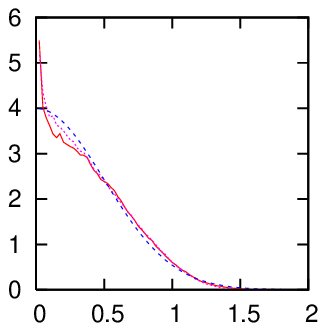}}
    \hspace{-0.3cm}
    {\includegraphics[height= 0.17\textwidth]{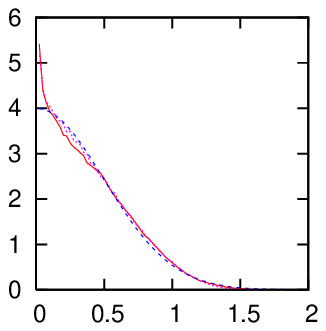}}\\
        {\includegraphics[height= 0.17\textwidth]{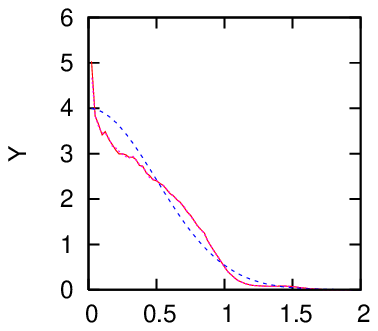}}
    \hspace{-0.3cm}
    {\includegraphics[height= 0.17\textwidth]{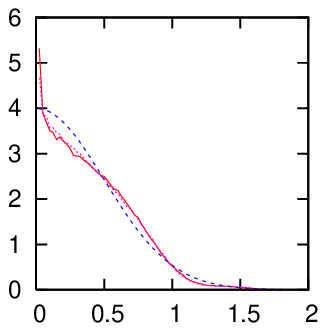}}
    \hspace{-0.3cm}
    {\includegraphics[height= 0.17\textwidth]{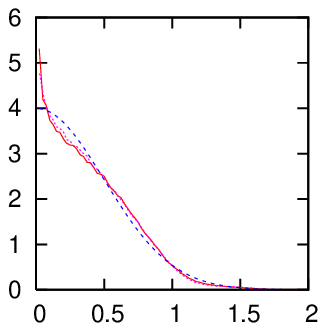}}
    \hspace{-0.3cm}
    {\includegraphics[height= 0.17\textwidth]{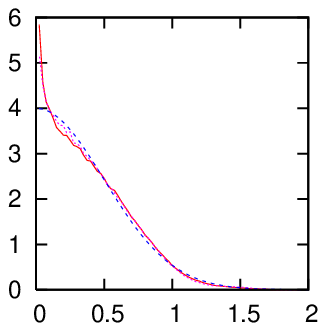}}
    \hspace{-0.3cm}
    {\includegraphics[height= 0.17\textwidth]{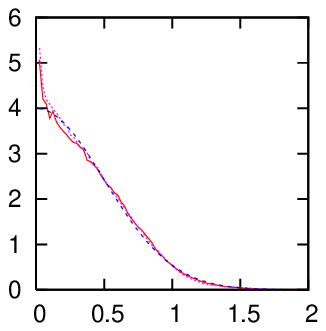}}\\
     \vspace{0.5cm}
    \hspace{-0.4cm}
    {\includegraphics[height= 0.17\textwidth]{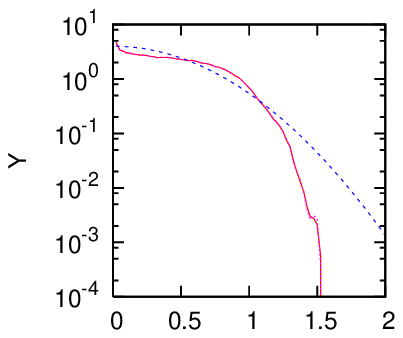}}
    \hspace{-0.4cm}
    {\includegraphics[height= 0.17\textwidth]{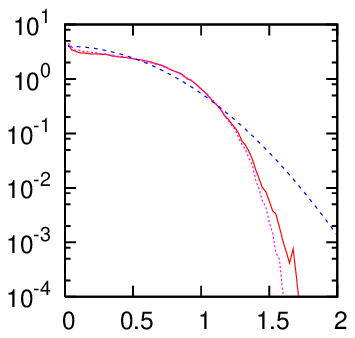}}
    \hspace{-0.4cm}
    {\includegraphics[height= 0.17\textwidth]{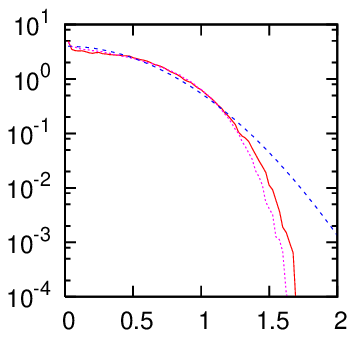}}
    \hspace{-0.4cm}
    {\includegraphics[height= 0.17\textwidth]{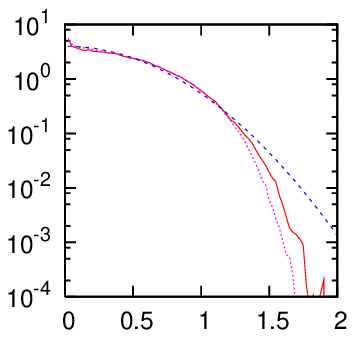}}
    \hspace{-0.4cm}
    {\includegraphics[height= 0.17\textwidth]{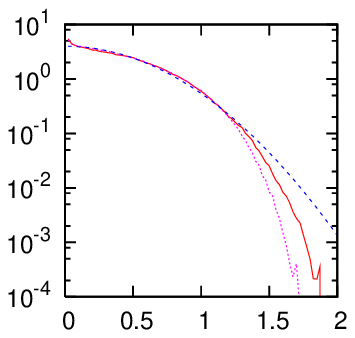}}\\
{\includegraphics[height= 0.19\textwidth]{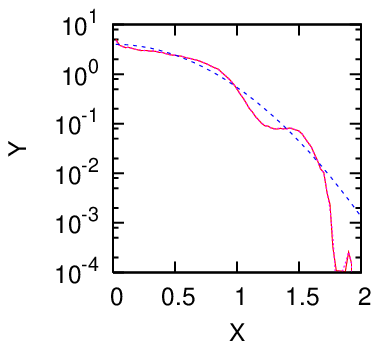}}
    \hspace{-0.8cm}
    {\includegraphics[height= 0.19\textwidth]{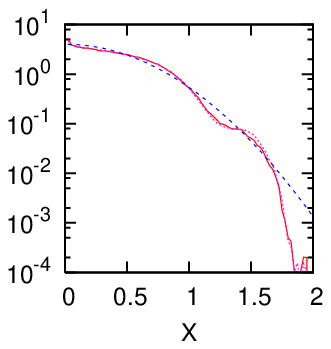}}
    \hspace{-0.8cm}
    {\includegraphics[height= 0.19\textwidth]{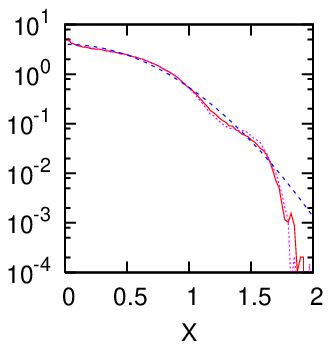}}
    \hspace{-0.8cm}
    {\includegraphics[height= 0.19\textwidth]{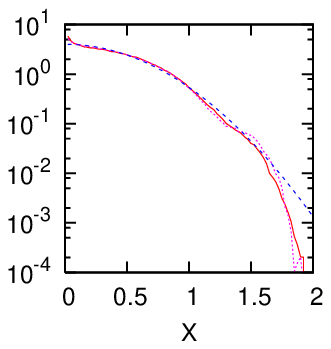}}
    \hspace{-0.8cm}
     {\includegraphics[height= 0.19\textwidth]{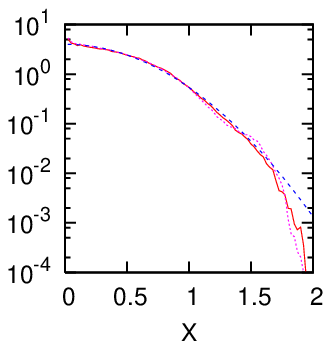}}
\end{center}
\caption{First row of plots:  evolution of the velocity
  pdf for $\mu_0=1$ and times $t=(20,1550,3100,4650,6200)\tau_{dyn}$.
 Second row of plots:  evolution of the velocity
  pdf for $\mu_0=1.7$ and times $t=(20,520,1030,1550,2060)\tau_{dyn}$.
The second block of plots are exactly the same but in log-linear scale. The plain red curve represents the simulations, the pink dotted one the theoretical prediction and the blue dashed curve the thermal equilibrium pdf \eqref{distri_MB}.}
\label{vel_all_theo}
\end{figure*}
In Fig.~\ref{vel_all_theo} we show the evolution of the full velocity
pdf for both the simulation and the theory. The first two rows of the
figure corresponds to the case $\mu_0=1$ and $\mu_0=1.7$
respectively. In the next two rows of the figure we reproduce the same
plots but in log--linear scale to appreciate the tails of the
distribution. We observe that the model predicts very well the
evolution of the velocity pdf for \emph{intermediate} values of the
velocities. For low velocities it predicts systematically a relaxation
\emph{faster} than the observed in the simulation, whereas for large
velocities it predicts systematically a relaxation slower than the one
observed in the simulations (in the latter case specially for the
$\mu_0=1.7$ system). We will discuss this discrepancy in the following
section.

\section{The validity of the Chandrasekhar approximation applied to inhomogeneous systems}
\label{sect-chandra}

\begin{figure}
  \begin{center}
    \psfrag{X}[][]{$y_0/x_0$}
    \psfrag{Y}[][]{$P(y_0/x_0)$}
    \psfrag{A}[][]{$\mu_0=1$}
    \psfrag{B}[][]{$\mu_0=1.7$}
    \includegraphics[height= 0.35\textwidth]{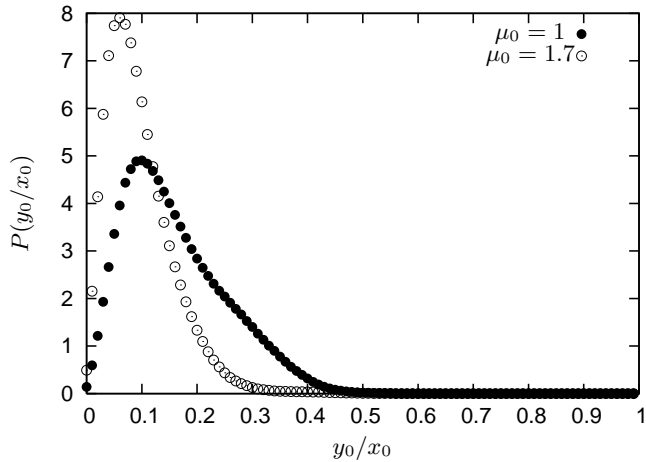}
\end{center}
\caption{Distribution of eccentricities $P(y_0/x_0)$ measured at $t=50\tau_{dyn}$ for both initial conditions.}
\label{fig-elip}
\end{figure}

It is possible to show that the result \eqref{v_perp} is the same one than the one obtained in
the spatial homogeneous case originally treated by Chandrasekhar
applied to gravity in $d=2$. In this study, it was considered
rectilinear trajectories with constant relative velocity $V$
(e.g. \cite{binney}), in which the distance of closest approach $y_0$
is the impact factor $b$. Then
\be
\label{chandra}
|\Delta
\bV_\perp|\simeq 2\int_{0}^{\infty} \frac{g\,b}{b^2+(Vt)^2}dt=\frac{g\pi}{V}.
\ee
The agreement between the results can be understood 
for two reasons:
\begin{enumerate}
\item trivially, in the limit $y_0/x_0\to0$, the unperturbed trajectories \eqref{elip} become rectilinear, and

\item  an excellent approximation to the integral \eqref{chandra} is obtained
taking $t\simeq b/V$ as upper cutoff, i.e., the collision is localized
in the same sense than the one discussed for the integral of
Eq.~\eqref{delta_vy}. 
\end{enumerate}
Therefore we can conclude, that when the relative orbits have large
ellipticity, the system can be treated as locally homogeneous and Eq.~\eqref{chandra} would be a good approximation. We have checked that it
is the case in our system, as it can be seen in
Fig.~\ref{fig-elip}. 
In this figure, we measure from the simulations
the value of $y_0/x_0$ for all the possible relative orbits
(i.e. $N(N-1)/2$ in total) at $t=50\tau_{dyn}$. We stress however that, as discussed above, it is not possible to average properly over velocities: the appropriate velocity pdf which must be used in Eqs.~\eqref{delta_vy} and \eqref{chandra} is not the velocity pdf but the  \emph{velocity pdf at the moment of the collision}. Having this idea in mind we obtain a very coherent picture to explain the results obtained in Fig.~\ref{vel_all_theo}:
\begin{itemize}
\item Particles with large velocity are very likely to be at the
  perigee of their orbit, i.e., the portion of the orbit in which the
  velocity is maximal. Hence, during the successive collisions, it is
  very probable that they would be in another portion of their orbit,
  with smaller velocity. Therefore, velocities at the moment of the collisions is systematically overestimated and, using Eq.~\eqref{delta_approx} (or
  Eq.~\eqref{coeff-expl-bessel}), the relaxation rate predicted by the
  Chandrasekhar approximation will be faster than the one which
  actually happens in the system.

\item The opposite occurs for low velocities: particles are more
  likely to be at the apogee of their orbit.  Therefore, the velocity
  in the moment of the collisions is systematically underestimated,
  and then, for the same reason than above, the Chandrasekhar
  approximation predicts a relaxation rate slower than the one which
  actually occurs.

\end{itemize}

The arguments presented above apply also in $d=3$, which may explain why
the original Chandrasekhar approach gives a good estimate of the
relaxation time in inhomogeneous systems, taking as maximal impact
parameter the size of the system (see e.g. \cite{farouki_94,diemand_04}). In particular:

\begin{itemize}

\item We expect that, in the same way than in the case studied here,
  the mean field potential would not change too much during the
  collisional relaxation process, which essentially makes the
  dynamical time $\tau_{dyn}$ invariant.

\item It has been shown numerically in $d=3$ that , for a wide set of
  initial conditions, the QSS presents also a central homogeneous
  region which decays rapidly to zero at larger scales
  \cite{roy+perez_2004}. Our hypothesis of
  Subsect.~\ref{mean-field-potential} would be therefore fulfilled.

\item All he arguments of Subsect.~\ref{comp-change-vel}  would also
  be true, and in particular the change of velocity due to one
  collision would have the same properties than Eq.~\eqref{delta_vy}, as we will show below.

\end{itemize}

Because collisions occurs also in a plane, we would have now
\bea
\label{delta_vy_3d}
\nonumber
|\Delta \bV_y|&\simeq& 2g \int_0^\frac{\pi}{2\omega}\frac{y_0\cos(\omega t)\,dt}{(x_0^2 \sin^2(\omega t) + y_0^2 \cos^2(\omega t))^{3/2}}\\
&=& \frac{2g}{\omega x_0 y_0}.
\eea
In the limit $y_0/x_0\to0$, and using Eq.~\eqref{v-approx}, we get the well-known result of Chandrasekhar  \cite{chandra_42}:
\be
\label{grav-3d}
|\Delta V_\perp|\simeq \frac{2g}{V b}.
\ee
Collisions are then ``local'', in the same manner than in the case
discussed in the paper, i.e., the change in velocity occurs in a
region of space of order of the impact factor. Analogously than
in the case treated in the paper it is difficult to estimate the
statistics of the relative velocities at the distance of closest
approach. However, the dependence of the change in velocity with the
impact factor is expected to be an excellent approximation. As
Eq.~\eqref{grav-3d} factorizes between a part which depends on the
velocity and another one on the impact factor $b$, even if we do an
error computing averages over velocities we would obtain the Coulomb
logarithm $\ln(R/b_{min})$ integrating over the allowed impact factors
($b_{min}$ is the minimal impact factor).  This explains why the
relaxation rate measured in simulations scales with the Coulomb
logarithm, as observed in simulations in gravitational systems in
$d=3$ \cite{farouki_82,smith_92,farouki_94,gabrielli_10,marcos_13}.

\section{Discussion}
\label{conclusion}
In this paper we have shown that using a ``minimal'' model --- based
on the Chandrasekhar approximation --- we can describe well the
evolution of the velocity distribution of a gravitational system in
$d=2$, for times from the formation of the QSS to thermal equilibrium. 
We have
derived an explicit kinetic equation neglecting collective effects, in which we
slightly adjust a single free parameter $r^*$. Comparing the
evolution of the velocity distribution observed in the simulation and
the one calculated with the model, we obtain a good agreement for all
times, from the formation of the QSS to thermal equilibrium.

We can conclude, as we anticipated in the Introduction, that the
Chandrasekhar (or local) approximation gives a reasonable description
of the collisional relaxation in this system. This is due to the fact
that many of the relative orbits of the particles which can be well
approximated by ellipses which large ellipticity, for which the
Chandrasekhar approximation is a good one. However, a systematic error
is made computing the diffusion coefficients, because the velocity of
the particles \emph{during the collisions} does not correspond in
general to the velocity of the particle at the moment in which we
sample the velocity pdf. Because of that, we have shown that we
overestimate systematically the relaxation rate of the particles with small
velocity and we underestimate systematically the relaxation rate of
particles with large velocities.

We have neglected possible resonances of the particles with the mean
field potential. We expect that they are not important, because particles have the same mass, which is a very different
situation of the decay of a single much massive particle inside a QSS
formed by much lighter ones, which can excite resonances (see
e.g. \cite{read_06}). Moreover, the actual potential in which
particles are moving is not harmonic but is close to the one of
Eq.~\eqref{pot_MB}: particles present highly precessing quasi-periodic
orbits, which are very unlikely to excite resonances by crossing the
system again and again following the same trajectory.

  On an other side, we do not observe numerically the scaling
  $\tau_{coll}\sim N^{1.35} \tau_{dyn}$ observed in
  \cite{teles_10}. This is is due to the fact that they use a
  simplified dynamics (polar symmetry is imposed along all the run and
  therefore particles conserve their initial angular momentum) appear
  not to describe properly the collisional dynamics of the real $d=2$
  system. A possible explanation of this discrepancy is that the model
  presented in \cite{teles_10} is not truly two-dimensional but quasi
  one-dimensional. It is known that one-dimensional models as the HMF
  can present striking scalings of the relaxation time with $N$, as
  pointed out in the introduction. Interestingly, however, the same
  group get, using the same simplified dynamics in $d=3$, the same
  scaling $\tau_{coll}\sim N \tau_{dyn}$ observed using full numerical
  simulations \cite{levin_etal_2008}. More investigation should be
  done to understand this discrepancy.

Some conclusions can be made about the maximal impact parameter which
has to be considered in the calculations. In the simulations we do not
observe any dependence of $r^*$ --- which is directly related with
the maximal impact parameters allowed --- with the number of particles
$N$. We can conclude then that the maximum impact parameter does not
depend on a scale related to the interparticle distance --- which
scales as $N^{-1/2}$--- but with the size of the system. Moreover, we
obtain an actual value of $r^*$ which corresponds to the size of the
homogeneous part of the system. This result is in agreement with
simulations performed in $d=3$ dimensions \cite{marcos_13} with
potential interactions $u(r)\sim 1/r^\gamma$, $\gamma \le 2$, in which
the maximal impact parameter to take in the Chandrasekhar
approximation was numerically estimated to be $1/3$ the size of the
system.

We can conclude that to obtain a better description of the collisional
relaxation, the use of action -- angle variables is
unavoidable. When performing the the calculation of
Eq.~\eqref{delta_vy} we are indeed using 
action -- angle variables, the parameters $x_0$ and
$y_0$ being proportional to the two actions of the system. A complete calculation using canonical perturbation theory is however much more involved.

\section*{Acknowledgments} 
I am very grateful to M.~ Joyce and Y.~Levin for many discussions
which made this work possible. I acknowledge for many useful
discussions J.~Barr\'e, C.~Nardini, R.~Pakter, F.~Peruani,
A.~C.~Ribeiro~Teixeira, T.~Teles and D.~Vincenzi. I warmly thank
M.~Courtney for her lecture of a previous version of the paper.
Numerical simulations have been performed at the cluster of the SIGAMM
hosted at ``Observatoire de C\^ote d'Azur'', Universit\'e de Nice --
Sophia Antipolis. This work was partly supported by the ANR
09-JCJC-009401 INTERLOP project and the CNPq PDS 158378/2012-1 grant.

\appendix
\section{Computation of the diffusion coefficients}
\label{derivation}
We define a
laboratory Cartesian system of coordinates with unit vectors $\hat
e_i$ ($i=1,2$), and another Cartesian system of coordinates $\hat
e'_i$, in which $\hat e'_1$ is in the direction of the initial
relative velocity. We have therefore 
\be
\Delta \bv= -|\Delta \bv_\parallel| \hat e'_1 + |\Delta \bv_\perp| \hat e'_2
\ee
The projection of the velocity in the $\hat e_i$ direction is then 
\be
\Delta v_i = -|\Delta \bv_\parallel| \hat e'_1\cdot\hat e_i  + |\Delta \bv_\perp| \hat e'_2\cdot\hat e_i.
\ee
Taking into account that, in average, collisions which will give
rise to a change of the perpendicular velocity are equally probable in
the $\hat e'_2$ direct in and in the direction opposite to it, we can
write
\bse
\label{diffusion_coeff}
\begin{align}
\Delta v_i  &= -|\Delta \bv_\parallel|  \frac{V_i}{V}\\
\Delta v_i \Delta v_j &=  |\Delta \bv_\perp|^2 \left(\delta_{ij}-\frac{V_iV_j}{V^2}\right),
\end{align}
\ese
where we have kept only the terms of $\mathcal O\left(g^2\right)$ and
used that $\hat e'_1\cdot\hat e_i= V_i/V$ and $(\hat e'_2\cdot\hat
e_i)(\hat e'_2\cdot\hat e_j)=\delta_{ij}-V_iV_j/V^2$. The diffusion coefficients are:
\bse
\begin{align}
D_{v_i}&=\frac{\langle \Delta v_i\rangle}{\Delta t}=-C\int d^2v' s(v') \frac{V_i}{V^3}\\
D_{v_i v_j} &=\frac{\langle\Delta v_i v_j\rangle}{\Delta t}=C\int d^2v' \frac{s(v')}{V}\left(\delta_{ij}-\frac{V_iV_j}{V^2}\right).
\end{align}
\ese

\begin{figure}
  \begin{center}
\psfrag{X}{v}
        {\includegraphics[height= 0.25\textwidth]{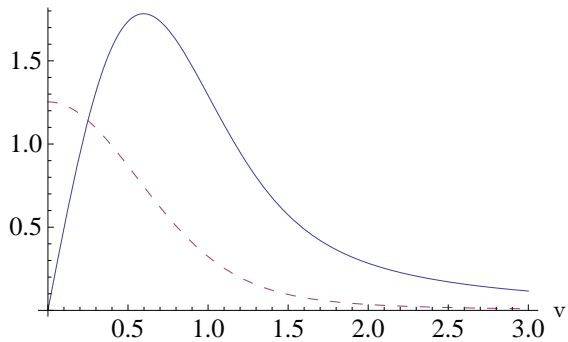}}
\end{center}
\caption{Plot of $q'(v)$ (straight line) and $p''(v)$ (dashed line) in function of $v$.\label{coefficients}}
\end{figure}

Introducing, as in the $d=3$ case, the Rosenbluth potential, we can write the diffusion coefficient using Eqs.~\eqref{coll} and \eqref{coeff}
\bse
\label{diff-rosen}
\begin{align}
D_{v_i}(v)=C\frac{\dd q(v)}{\dd v_i}\\
D_{v_i v_j}(v) =C\frac{\dd^2 p(v)}{\dd v_i \dd v_j},
\end{align}
\ese
where
\bse
\begin{align}
q(v) &= \int d^2v' \frac{s(v')}{|\bv-\bv'|}\\
p(v) &= \int d^2v' s(v')|\bv-\bv'|,
\end{align}
\ese
where we have assumed that the velocity pdf is isotropic. Using that the Rosenbluth potentials are isotropic we can simplify Eqs.~\eqref{diff-rosen} using that
\bse
\label{rosen-polar}
\begin{align}
\frac{\dd q(v)}{\dd v_i}&=\frac{v_i}{v}q'(v)\\
\frac{\dd^2 p(v)}{\dd v_i \dd v_j}&=\frac{v_iv_j}{v^2}\left(p''(v)-\frac{p'(v)}{v}\right)+\delta_{ij}\frac{p'(v)}{v},
\end{align}
\ese
where the prime denotes derivative with respect to $v$. In Fig.~\ref{coefficients} we plot $q'(v)$ and $p''(v)$, which gives of $D_{v_i}(v)$ and $D_{v_iv_j}(v)$ respectively.


\end{document}